# Traffic Convexity Aware Cellular Networks: A Vehicular Heavy User Perspective

*Taehyoung Shim, Jihong Park, Seung-Woo Ko, Seong-Lyun Kim,*

*School of Electrical and Electronic Engineering*
*Yonsei University, Seoul, Korea*
*e-mail: {teishim, jhpark.james, swko, slkim}@ramo.yonsei.ac.kr*

*Beom Hee Lee, and Jin Gu Choi*

*LG Electronics, Korea*
*e-mail: {beomhee.lee, jingu.choi}@lge.com*

***Abstract*—Rampant mobile traffic increase in modern cellular networks is mostly caused by large-sized multimedia contents. Recent advancements in smart devices as well as radio access technologies promote the consumption of bulky content, even for people in moving vehicles, referred to as vehicular heavy users. In this article the emergence of vehicular heavy user traffic is observed by field experiments conducted in 2012 and 2015 in Seoul, Korea. The experiments reveal that such traffic is becoming dominant, captured by the 8.62 times increase in vehicular heavy user traffic while the total traffic increased 3.04 times. To resolve this so-called vehicular heavy user problem (VHP), we propose a cell association algorithm that exploits user demand diversity for different velocities. This user traffic pattern is discovered first by our field trials, which is convex-shaped over velocity, i.e. walking user traffic is less than stationary or vehicular user traffic. As the VHP becomes severe, numerical evaluation verifies the proposed user convexity aware association outperforms a well-known load balancing association in practice, cell range expansion (CRE). In addition to the cell association, several complementary techniques are suggested in line with the technical trend toward 5G.**

# I. INTRODUCTION

The nomadicity of users has long been regarded as the traffic pattern of moving users whose data consumption is mostly concentrated on their stationary states. The recent advancement in cellular networks along with the smart device proliferation, however, promotes the data consumption during the in-transit states. This emerging trend may considerably alter the overall traffic pattern if the dominant traffic volume originates from in-vehicle passengers who are able to consume even a large-sized multimedia content. We thus redefine such nomadic users as *vehicular heavy* users, and revisit their impact on the overall traffic pattern.

From a network design perspective, vehicular heavy users consuming bulky multimedia contents may become a critical challenge for the next generation cellular networks. Indeed, their huge demand for radio resource per se is a strain on network operators. In addition, frequent handovers incurred by quickly moving vehicles force the spectral efficiencies to take a nosedive. Unabated vehicular heavy user traffic would therefore bring about the rapid depletion of available radio resource, degrading network capacity.

In this article we foresee that such a vehicular heavy user problem (VHP) is imminent based on a litmus test conducted in Seoul, Korea–a cutting-edge cellular technology test bed–where the world's first tri-band carrier aggregation has already been applied with 63.4% LTE penetration rate since January, 2015 [1]. As a remedy for the upcoming VHP, we focus on a convex-shaped user traffic pattern first proved by the field experiment, and propose the user convexity aware cell association accordingly.

## CONVEX TRAFFIC OVER VELOCITY

Users riding a bus are able to watch large-sized video contents [2] as in stationary users whereas walking users cannot concentrate on these types of contents [3]. This may lead to convex-shaped user traffic over velocity. Our field experiments conducted in 2012 and 2015 confirm such a traffic pattern conjecture. Furthermore, the results reveal that the vehicular heavy user traffic proportion increase is 1.29 times larger than the walking users' during the period. Such an increasing heavy-to-light traffic discrepancy motivates us to exploit the convex traffic pattern for a cellular network design so as to resolve the VHP.

### CONVEX TRAFFIC AWARE NETWORK DESIGN

We consider a two-tier downlink cellular network comprising small and macro cells, and design the network in a way that improves average rate while guaranteeing each user's minimum rate requirement. Recalling the VHP, our prime concern is vehicular heavy users' minimum rate satisfaction lest it become a bottleneck. In this respect, we propose a cell association algorithm that increases the vehicular user rate in return for sacrificing the walking user rate. The lowest traffic volume of walking users, observed from the convex traffic pattern, allows them to put up with such rate degradation while improving average rate. In view of its exploiting velocity-specific user demand diversity at cell associations, our proposed network design is in line with the state-of-the-art service-oriented architecture [4] that flexibly controls core-to-wireless network operations with service information, i.e. user demand. Its implementation in practice is to be enabled by 5G cellular technologies such as cloud radio access network (C-RAN) [5] and software-defined network (SDN) [4], respectively allowing centralized management and layer-transcending optimization in a flexible manner.

## II. USER TRAFFIC PATTERN OVER VELOCITY

The recent rise of smart devices and advancements in cellular technologies even allows moving users to enjoy multimedia contents without technical limitation. This motivates us to revisit the mobile data traffic pattern with respect to user velocity. In this section we investigate traffic pattern over velocity through field experiments, which forecasts that the VHP is imminent.

### EXPERIMENTAL SETUP AND PROCEDURES

Our objective is to establish a downlink data traffic tendency along with user velocity. For this purpose, we conducted field experiments twice in Seoul, Korea during: (i) May–June in 2012 and (ii) March in 2015 when LTE penetration rates respectively are 11.2% and 63.4% [1]. Under the environment summarized in Table I, total 152 participants are recruited. They installed an Android application (LifeMap [6] in 2012, My Data Manager[1] with Moves[2] in 2015) that measures their downlink data usage and locations at intervals of 5 minutes during the experimental periods. When each user was assumed to move linearly in-between two consecutive locations, velocity is calculated by dividing the corresponding distance by the given time interval. The measured data usage and velocity information was then collected and processed as follows.

[1] *My Data Manager* is available at: *http://www.mydatamanagerapp.com.*
[2] *Moves* is available at: *https://www.moves-app.com.*

**User mobility state definition**–Users are divided into three mobility states: *stationary users* do not change their positions, interpreted as people working in an office or sitting in a café; *walking users* are strolling or jogging people; and *vehicular users* are in-vehicle passengers, hereafter identically treated as vehicular heavy users for notational brevity.

**User mobility state separation**–Vehicular users were recognized when their velocities exceeded 10 km/h, rush hour vehicle velocity in Seoul, Korea. Along with the measured location recording, such a detection had an interval of 5 minutes. Note that this interval corresponds with average time duration of unit linear movement during a single travel by vehicle [7], and is thus sufficiently short for vehicular user recognition. On the other hand, walking and stationary users were recognized by exploiting GPS, accelerometer, and digital compass information from the participants' smart devices. The installed measuring tool kept a watch on the sensors at intervals of several seconds. Detecting a sudden change in the sensor information led to walking/stationary user recognition. Figure 1 provides a typical user example of our experimental data collection.

**Measurement aggregation of multiple users**–In consequence of the said measurement and calculations, every user generated the pairs of data usage and velocity in series during the experimental period. The series of data were averaged over time respectively for three different mobility states. This yields stationary/walking/vehicular state traffic volumes in units of MB/day for each user. The time-averaged traffic volumes of multiple users were then averaged over users for different mobility states, resulting in an arbitrary user's average stationary/walking/vehicular state traffic volumes in units of MB/day.

Note that the measuring tool in 2012 was inevitably replaced in the 2015 experiment since the old one was no longer compatible with the participant's devices in 2015. Nevertheless, its impact is negligible during our data collection, and therefore the traffic pattern comparison between 2012 and 2015 is valid. Further justification of this issue is deferred to a supplementary document that incorporates the measured raw data in

2012 and 2015[3].

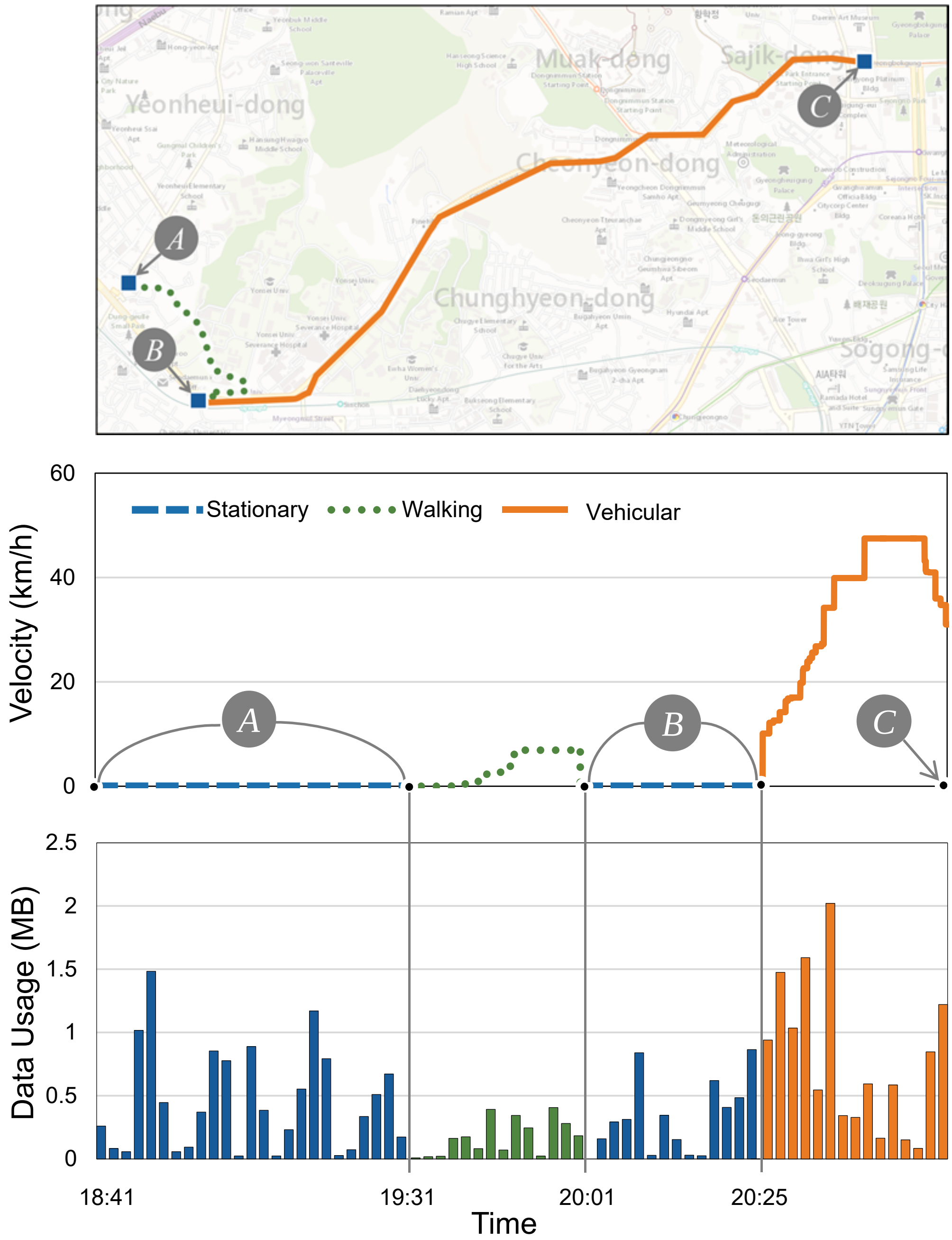

**Figure 1.** An experimental example of a movement trace, velocity, and the corresponding data usage of a typical user in Seoul, Korea. The user remained stationary (blue) at *A* for a while, and walked (green) to *B*. After a short stay at *B*, he moved to *C* by a vehicle (orange). The corresponding data usage indicates that the user in stationary and vehicular mobility states consume more data than during the walking state.

| *Year* | *2012* | *2015* |
|---|---|---|

[3] The measured raw data in 2012 and 2015 for further reference are available at: *http://tiny.cc/convexityexperiment*.

| | | |
|---|---|---|
| *Number of participants* | 82<br>(male: 61, female: 21) | 70<br>(male: 54, female: 16) |
| *Age of participants* | avg. 25.9 | avg. 27.8 |
| *Experimental period* | avg. 1.28 weeks | avg. 1.13 weeks |
| *Radio access technology* | WCDMA / LTE | WCDMA / LTE / LTE-A |
| *Measurement applications* | LifeMap | My Data Manager, Moves |
| *Velocity* | *Walking*: up to 10 km/h (avg. 2.58 km/h)<br>*Vehicular:* above 10 km/h (avg. 31.9 km/h) | |

**Table 1.** Experimental environments in 2012 and 2015.

## KEY OBSERVATIONS

Our field experiment results are summarized in Figure 2. Comparing the results of 2012 and 2015, we provide three key observations as follows.

1. **Total traffic volume increase**–The average daily downlink data usage in 2015 increased by 3.04 times from 2012 (47.66 → 145.05 MB/day). This explosive traffic growth results from advancements in LTE/LTE-A as well as prevailing smart devices as observed also by [8].
2. **Moving (walking/vehicular) user traffic proportion increase**–Moving user traffic proportion of the total traffic volume increased by 2.64 times (14.74 → 38.93%). Not only was there a rapid growth in moving user traffic, but also the slow growth of stationary user traffic. Moving users are comfortable with consuming multimedia contents thanks to high performance smart devices [8] and mobility supporting technologies such as seamless handover schemes [9], leading to increased moving user traffic. Stationary users, on the other hand, frequently offload their traffic to Wi-Fi [10], resulting in the slow growth of stationary user traffic.
3. **Vehicular heavy user traffic proportion increase**– Vehicular user traffic proportion increased by 2.83 times (10.35 → 29.29%). The value corresponds to the 8.62 (4.93 → 42.48 MB/day) times traffic volume growth, which is much larger than the 3.04 times total traffic volume growth during the same period. Walking user traffic proportion, on the other hand, increased only by 2.2 times (4.39 → 9.65%). As a consequence, the traffic proportion increasing rate of vehicular users is 1.29 times faster than that of walking users. This reinforces a convex-shaped traffic pattern, also captured by a typical user's example in Figure 1. Such a pattern emerges from the fact that vehicular users are able to concentrate on bulky multimedia contents whereas walking users are not [3].

The above experimental results underpin our prediction that vehicular heavy user traffic–1/3 of the entire traffic volume as of 2015–will no longer be negligible. This observation in Seoul, Korea, having the world's most advanced cellular infrastructure [1, 8], may allow us to give a glimpse of the global traffic pattern in near future. Beyond our observation, we further anticipate that the VHP will become even more challenging when considering the recent cellular base station densification trend toward ultra-dense networks [11]. It may incur more frequent handovers, degrading the spectral efficiency of vehicular users. In addition to this trend, the VHP may also intensify along with the upcoming services that promote vehicular user traffic consumption such as vehicle-to-infrastructure and high-speed railway communications as well as driverless cars. In the following section, we thus propose a novel network design approach to cope with the VHP.

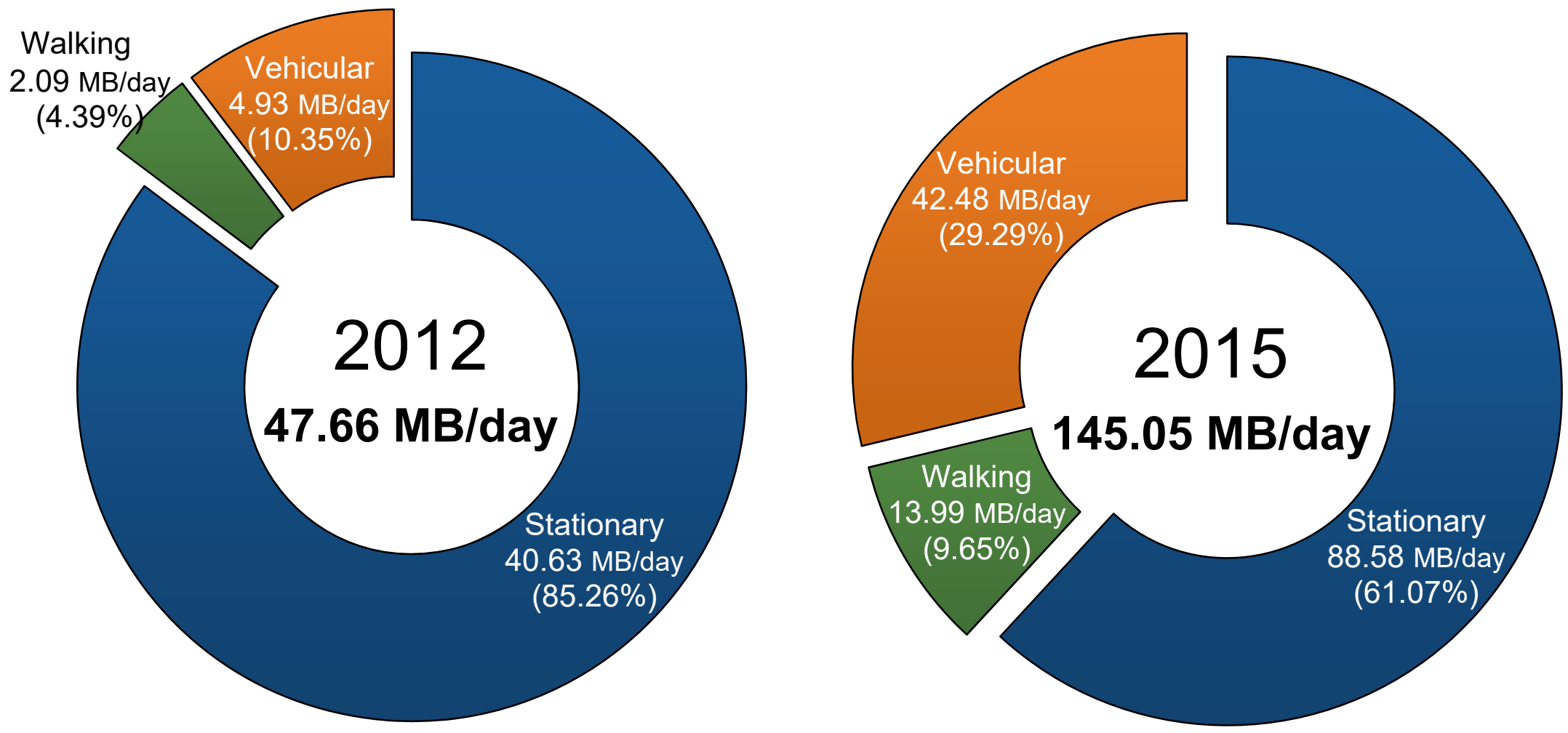

**Figure 2.** Total traffic volumes and traffic volume proportions of stationary, walking, and vehicular users in 2012 and 2015. Vehicular heavy user traffic volume increased by 8.62 times from 2012 to 2015 (4.93 → 42.48 MB/day) while total traffic volume only increased by 3.04 times (47.66 → 145.05 MB/day), leading to the 2.83 times (10.35 → 29.29%) vehicular heavy user traffic proportion increase.

## III. User Convexity Aware Network Design

This section introduces a novel cell association algorithm to combat the VHP by utilizing user demand diversity over velocity, observed from our field experiments. The scheme gives association priorities to vehicular and stationary users, and not to walking users, so as to increase the vehicular user rate as well as the average rate of aggregate users. Since the association priority determination is vital to the proposed algorithm, we define a metric *user convexity* that quantifies the difficulty of the VHP and thereby determines the extent to which the associations should be prioritized. Such a user convexity aware association pertinent to the VHP is numerically evaluated, and its implementation under practical cellular architecture is described in the following subsections.

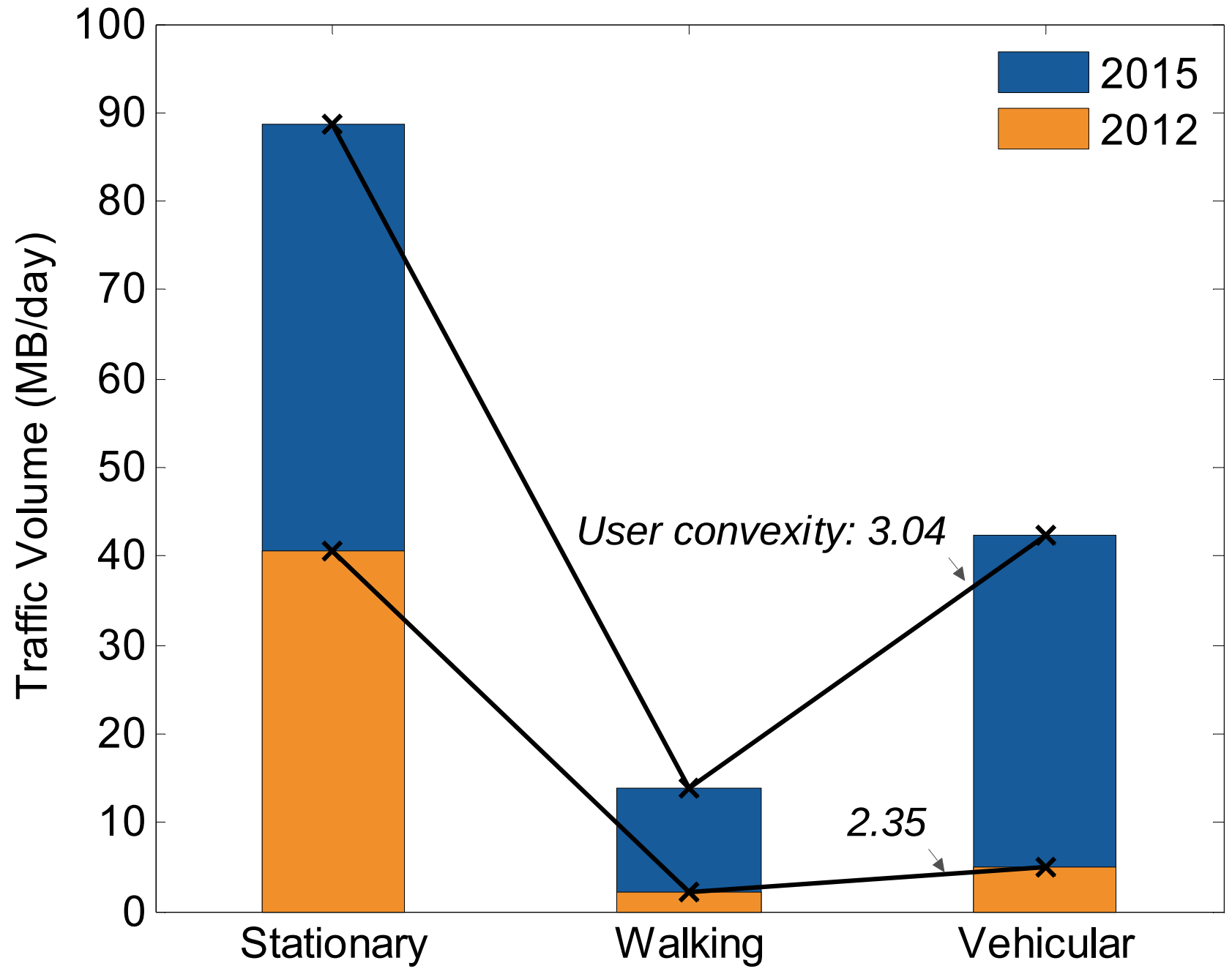

**Figure 3.** Traffic volume over velocity and its corresponding user convexity. Stationary and vehicular user traffic volumes are larger than the walking user volume. The difference increased from 2012 to 2015, measured by 1.29 times (user convexity: 2.35 → 3.04) increase in user convexity defined as the traffic volume ratio of vehicular users to walking users.

## USER CONVEXITY

User convexity is defined as the traffic volume ratio of vehicular users to walking users, geometrically incorporating the convexity of traffic volume curve over velocity. This value indicates the difficulty of the VHP. As an example, high user convexity corresponds to severe VHP due to heavy traffic generating vehicular users as well as their large traffic proportion of the entire traffic volume. Figure 3 represents our field experimental results with user convexity. The value in 2015 increased by 1.29 times from 2012 (2.35 → 3.04). This emphasizes the necessity of designing a user convexity aware network in order to combat the VHP.

## USER CONVEXITY AWARE CELL ASSOCIATION

In a two-tier cellular network comprising macro and small cells, we propose a cell association algorithm that facilitates the VHP mitigation by assisting vehicular heavy users in return for putting additional burden on other users. To minimize the inevitable damage, the victims to the algorithm are walking users whose traffic volume is the lowest. The scheme associates them with more congested cells in order to vacate frequency resource for vehicular users. User convexity at this point plays a key role to determine the association bias achieving the required resource to be vacated.

To be more specific, consider rate coverage [12] defined as the probability that each user state's average rate exceeds its expected traffic volume. On the basis of either macro or small cell association providing maximum received power, the proposed algorithm adjusts small cell association bias so that the rate coverage averaged over users is maximized while guaranteeing at least a target rate coverage for each user mobility state. In this problem, the minimum rate coverage constraint of vehicular users is likely to be a major bottleneck due to the VHP. The proposed scheme thus tries to satisfy the vehicular user's minimum rate coverage constraint with the highest priority, and then increase the average rate coverage as much as possible.

Such a goal is achieved via the following three-stage operations[4].

**Stage 1. Stationary user bias**– Increase small cell association bias of stationary users for their sole purpose of rate coverage maximization by neglecting walking and vehicular users. This operation improves average rate coverage by prioritizing the biasing factor decision of stationary users whose traffic volume is the majority.

[4] Mathematical descriptions of convexity aware cell association and its operations for different user convexity and traffic volume are deferred to: *http://tiny.cc/convexityalgorithm*.

**Stage 2. Walking user bias**– Increase small cell association bias of walking users in order to vacate macro cell resources for vehicular users (to be associated with macro cells in the next stage). The small cell biasing factor increases with user convexity. It implies more severe VHP requires more macro cell resources via migrating more walking users to small cells.

**Stage 3. Vehicular user bias**– Determine small cell association bias of vehicular users so as to maximize their own rate coverage under the given biasing factors of stationary and walking users in Stages 1 and 2. The optimal biasing factor of vehicular users is likely to associate a large number of vehicular users with macro cells to avoid frequent handovers due to fast velocity. At this point, the procured macro cell resources–by sacrificing walking users according to user convexity in Stage 2–are utilized in order to reduce macro cell congestion, mitigating the VHP.

Compared to preceding works, it is remarkable that our cell association algorithm counteracts traffic demand diversity for different velocities, captured by user convexity. The key operation is intentionally associating walking users to small cells (Stage 2) in order to associate vehicular users with the vacated macro cells (Stage 3). This VHP solution is motivated and determined by user convexity, i.e. walking users consume less traffic than vehicular users. In addition, the frequent handovers of vehicular users are mitigated by means of their macro cell associations. Furthermore, severe per-cell congestion is reduced via adjusting small cell association bias for load balancing of stationary users (Stage 1), corresponding with cell range expansion (CRE) [12] in practice. As a consequence, the proposed convexity aware association copes with user demand diversity, handover loss, and load balancing. To the best of our knowledge, this is the first work that considers these three factors simultaneously.

## NUMERICAL EVALUATION

For a given total traffic volume, we numerically evaluate the average rate coverage of user convexity aware association along with user convexity increase. For different total traffic volumes, we additionally derive the required bandwidth to achieve the minimum rate coverage requirements.

For user convexity increase, our proposed algorithm makes vehicular users more associated with macro cells, but walking users more with small cells. In so doing, procured macro cell resource for vehicular users mitigates the VHP. Figure 4 illustrates its impact where total traffic is given as 145.05 MB/day that corresponds to our results in 2015. The user group ratios are assumed as our experimental investigation; the group ratios of stationary, walking, and vehicular users are respectively 89.72%, 4.70% and 5.58%. For given 10 MHz downlink bandwidth, our proposed association (thick solid blue) outperforms CRE (thin solid orange), and the gap becomes large along with the user convexity increase. Such a rate coverage of the proposed association algorithm accords with at least 98% of the full-search performance (dotted black) for all user convexity values. In addition, when the minimum rate coverage thresholds are identically fixed at 0.8 for all users, the proposed scheme guarantees the minimum requirements for higher user convexity values, as opposed to CRE, captured by their vertical thresholds respectively at user convexity values of 5.2 and 7.1. These results indicate our proposed association is useful for severe VHP, i.e. high user convexity.

For total traffic volume increase, the proposed algorithm makes all users more associated with small cell as in CRE. Vehicular users, in distinction from other users, become less likely to associate with small cells to avoid frequent handovers. Its impact is visualized in Figure 5 via the required bandwidth to satisfy the minimum rate coverage requirements of users. The result validates our assertion that convexity aware association requires less bandwidth than CRE. It also shows that the required bandwidth gap between our proposed scheme and CRE becomes large when total traffic is doubled.

Such results show that convexity aware cell association is superior in the capabilities of the VHP mitigation as well as total traffic increase, the forthcoming major threats expected by our field experiments.

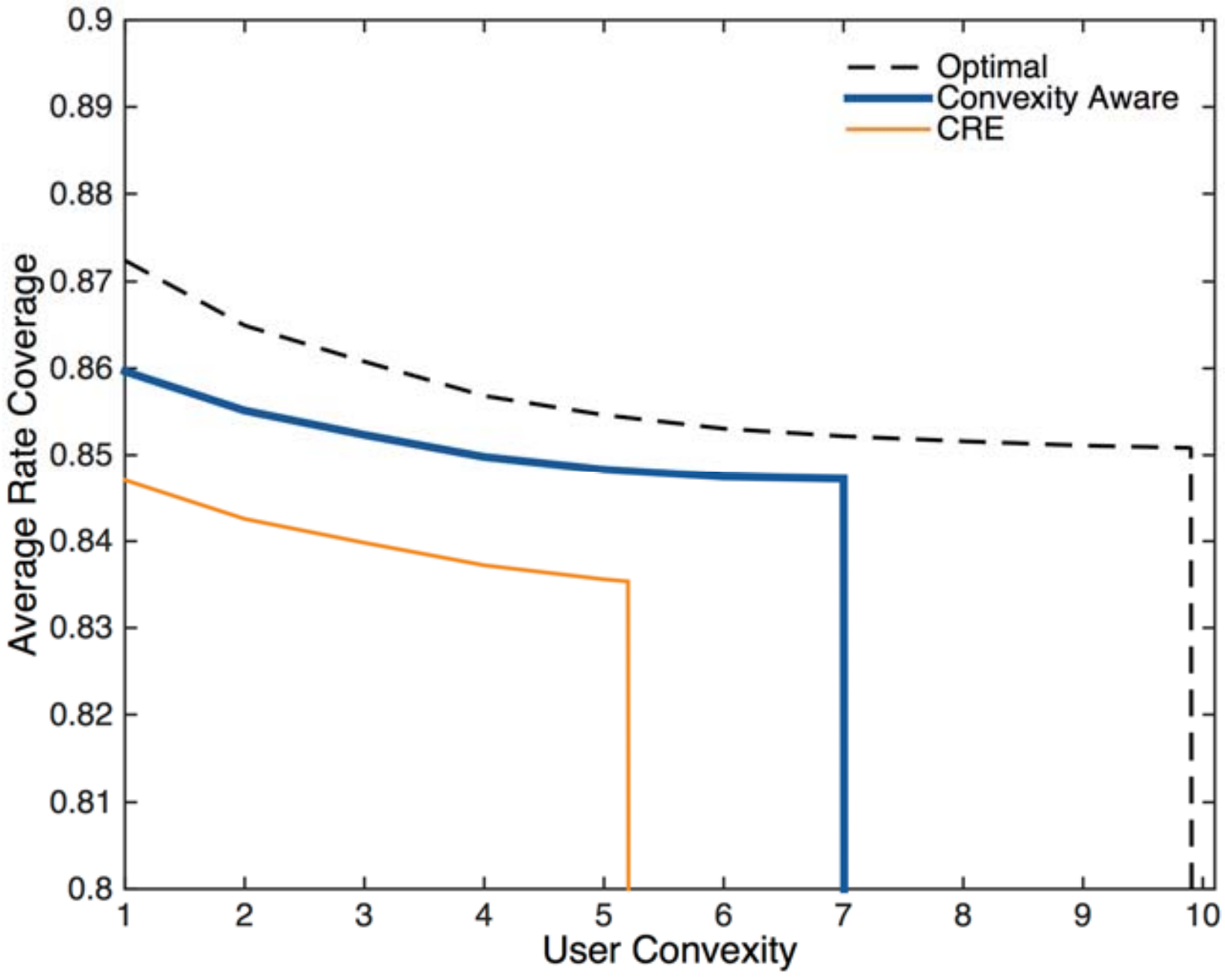

**Figure 4.** Average rate coverage with respect to user convexity. User convexity aware association provides higher average rate coverage than CRE, close to the optimal performance achieved by full search (total traffic = 145.05 MB/day, bandwidth = 10 MHz, rate coverage threshold = 0.8, user density: [stationary, walking, vehicular] = $[0.8972, 0.0470, 0.0558]/m^2$).

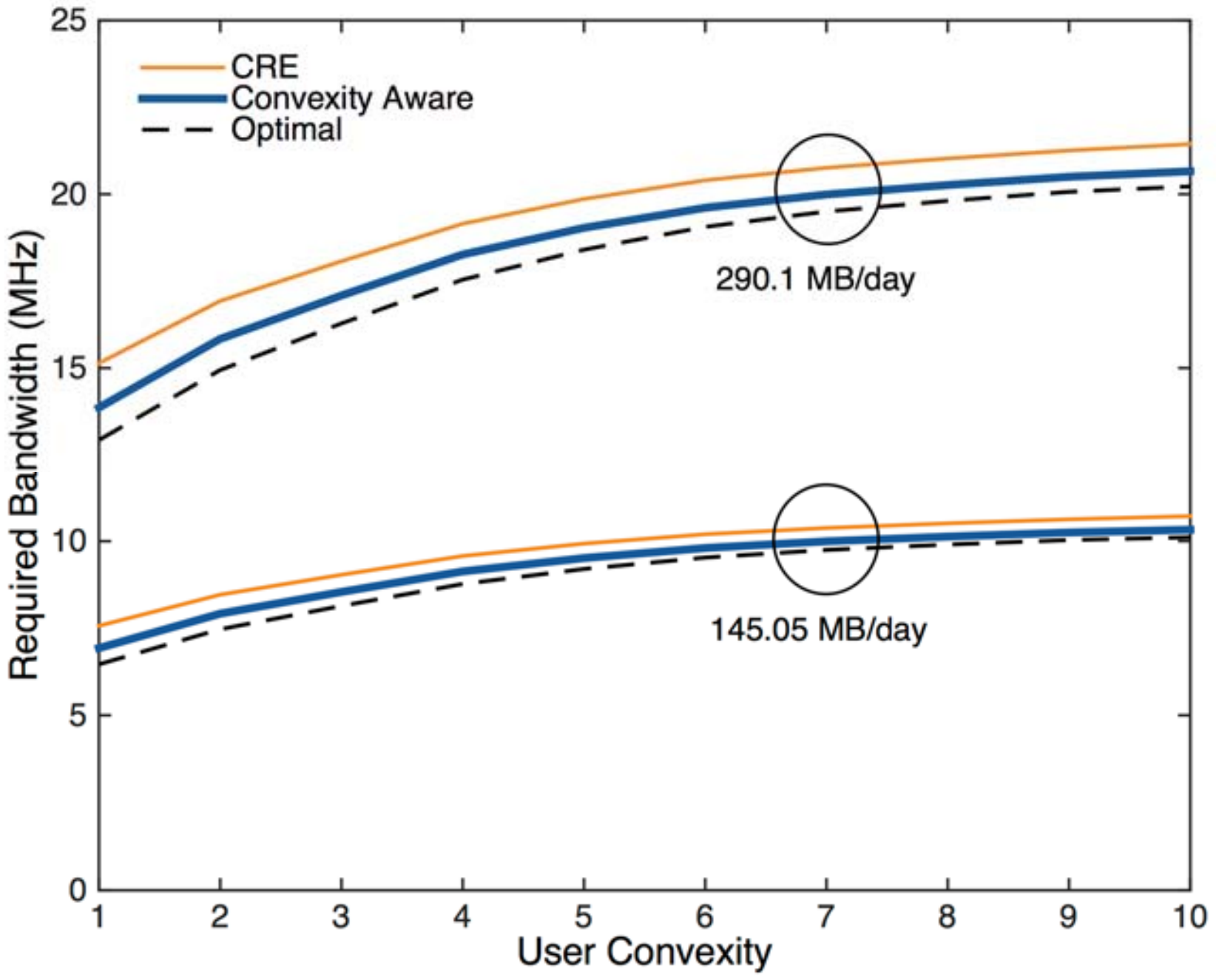

**Figure 5.** Required bandwidth to guarantee the minimum rate coverage requirements of users. Total traffic increase twice from 145.05 to 290.1 MB/day results in the larger required bandwidth gap between user convexity aware association and CRE (rate coverage threshold = 0.8, user density: [stationary, walking, vehicular] = $[0.8972, 0.0470, 0.0558]/m^2$).

### IMPLEMENTATION UNDER PRACTICAL ARCHITECTURE

The proposed user convexity aware cell association can be implemented under the sate-of-the-art cellular architecture in practice. C-RAN [5] in LTE core architecture enables the association via acquiring user request data amount and user velocity information. The information is by default collected by base stations. A base station receives its associated user's data request amount, and sends it to C-RAN. In a similar manner, the base station collects user location information comprising the desired signal timing advance, measured reference signal strength, and/or a Global Navigation Satellite System (GNSS) signal. This location information is reported to Evolved Serving Mobile Location Center (E-SMLC) in C-RAN at intervals ranging from 1 to 64 seconds [13]. C-RAN then calculates user velocities based on the reported information, and combines them with the corresponding data request amounts. The resultant information is used to determine the user convexity of the network and its matching association bias for the user groups having different velocities.

## IV. CONSIDERATIONS AND FUTURE CHALLENGES

The low volume of walking user traffic plays a key role in our proposed user convexity aware association. Emerging wearable and internet-of-things devices however may increase walking user traffic, leading to a weakening of the proposed algorithm's performance. Moreover, wirelessly powered [14] devices in the future will make walking users free from battery depletion, and hence consume more traffic, which may further degrade the proposed association. It is therefore necessary to keep an eye on the traffic pattern over velocity as in our field experiments for long-term period. For the preparation of such scenarios where our proposed scheme is hardly viable, complementary VHP solutions from different perspectives are suggested, opening interesting avenues for future research as follows.

**Handover reduction via dual connections**–The high mobility of vehicular users by itself contributes to the VHP due to its frequent handovers downgrading spectral efficiency. Dual connectivity [15], establishing both macro and small cell connections respectively for control signaling and data downloading, can be a possible solution. It would reduce the number of handovers, which would be conductive to alleviating the VHP.

**In-vehicle communication capability enhancement**–The rooftop transceiver installation of vehicles as a relay, a moving cell [16], enables a group-wise handover of vehicular users, and allows users to be free from vehicle penetration loss, increasing the ability to cope with the VHP. In addition, the promising massive antenna array equipment in 5G by utilizing the space in vehicles may further boost the capability.

**Caching at predictive vehicular users**–Prefetching data to the stationary or walking users who are expected to become vehicular users shortly afterwards reduces vehicular user traffic. User context based movement prediction enables such data caching. The context of those predictive vehicular users is in accord with, for instance, walking users around a parking lot or stationary users at a bus stop. Intelligent sensors in smart and wearable devices will aid in the prediction, and make this caching scenario viable in the near future.

## V. Conclusions

This article presents a two-fold experimental discovery of cellular user traffic in conjunction with velocity. The first compelling evidence is the rapid growth of vehicular heavy user traffic, i.e. the VHP. The second noticeable observation is the low traffic volume of walking users, yielding a convex traffic pattern. To circumvent the difficulty of the VHP, a biased cell association algorithm is proposed, which sacrifices walking user associations for increasing vehicular user rate. Such an algorithm relieves the VHP regardless of giving walking user associations less priority thanks to the convex traffic pattern. From the opposite angle, the proposed algorithm's dependency on the convex traffic pattern however can be a major drawback. Its follow-up experiments by periods thus will be necessary for years ahead. Furthermore, other complementary techniques pertinent to the VHP should be required. Concerning the recent technical trends toward 5G, dual connectivity, massive antenna array equipped moving cells, and predictive caching would be such candidates, and also could be the further extension to this work.

## Acknowledgement

The authors warmly thank Editor and anonymous Reviewers for the valuable comments enabling us to substantially improve the earlier version of this article. This research has been supported by LG Electronics Co., Ltd.